\documentclass[journal]{vgtc}                
\ifpdf
  \pdfoutput=1\relax                   
  \usepackage{graphicx}                
  \DeclareGraphicsExtensions{.pdf,.png,.jpg,.jpeg} 
\else
  \usepackage{graphicx}                
  \DeclareGraphicsExtensions{.eps}     
\fi%

\graphicspath{{figures/}{pictures/}{images/}{./}} 

\usepackage{microtype}                 
\PassOptionsToPackage{warn}{textcomp}  
\usepackage{textcomp}                  
\usepackage{mathptmx}                  
\usepackage{times}                     
\usepackage{cite}                      
\usepackage{tabu}                      
\usepackage{booktabs}                  

\usepackage[flushleft]{threeparttable}
\usepackage{xcolor}
\usepackage{paralist}
\usepackage{amssymb}
\usepackage{placeins}
\usepackage{scrextend}
\usepackage{stfloats}
\usepackage{tikz}

\DeclareMathAlphabet{\mathcal}{OMS}{cmsy}{m}{n}

\usepackage[draft=true]{minted}
\usepackage[hyphens]{url}
\usepackage[hidelinks]{hyperref}
\hypersetup{breaklinks=true}

\usepackage{multirow}

\newenvironment{rev}[0]{%
    \leavevmode\color{black}\ignorespaces
}{}

\newcounter{cIndex}

\newcommand{\criticism}[1]{{\stepcounter{cIndex}\fontfamily{lmss}\selectfont{\textbf{C\thecIndex}:~#1.}}}

\newcommand{\ci}[1]{{\fontfamily{lmss}\selectfont{C#1}}}

\usepackage{xspace,xpunctuate}
\newcommand{\ie}{{i.e.,}\xspace}
\newcommand{\eg}{{e.g.,}\xspace}
\newcommand{\ea}{{et~al\xperiod}\xspace}

\usepackage{amsmath}

\newcommand{\np}{24}
\newcommand{\nva}{47}
\newcommand{\ncrit}{257}
\newcommand{\npsecond}{17}
\newcommand{\nc}{36}

\definecolor{lGrey}{RGB}{245, 245, 247}
\definecolor{dGrey}{RGB}{210, 210, 210}

\definecolor{dRed}{RGB}{226, 127, 145}
\definecolor{lRed}{RGB}{255, 224, 224}

\definecolor{dRL}{RGB}{232, 169, 158}
\definecolor{lRL}{RGB}{255, 223, 217}

\definecolor{dPro}{RGB}{255, 239, 200}
\definecolor{lPro}{RGB}{253, 250, 227}

\definecolor{dData}{RGB}{202, 219, 191}
\definecolor{lData}{RGB}{246, 250, 241}

\definecolor{dVis}{RGB}{114, 194, 186}
\definecolor{lVis}{RGB}{220, 242, 240}

\definecolor{dSys}{RGB}{124, 173, 230}
\definecolor{lSys}{RGB}{225, 244, 253}

\definecolor{dEva}{RGB}{145, 142, 173}
\definecolor{lEva}{RGB}{222, 220, 245}

\definecolor{dDis}{RGB}{210, 210, 210}
\definecolor{lDis}{RGB}{245, 245, 247}

\definecolor{dOthers}{RGB}{210, 210, 210}
\definecolor{lOthers}{RGB}{245, 245, 247}

\newcommand{\crit}[3]{\vspace{2px} \noindent\begin{tikzpicture}
\draw [fill={#1},draw=none] (0,0) rectangle (0.2, 0.6) node[] {};
\draw [fill={#2},draw=none] (0.2,0) rectangle (8.75, 0.6) node[pos=.5] {\parbox{8.2cm}{\criticism{#3}}};
\end{tikzpicture} \vspace{2px} \\ \indent}

\usepackage{tikz}
\definecolor{nodeBG}{RGB}{29,40,97}

\usepackage{amsmath}

\usepackage{array}

\usepackage[ruled,vlined]{algorithm2e}
\SetKwComment{Comment}{$\triangleright$\ }{}

\onlineid{1035}

\vgtccategory{Research}
\vgtcpapertype{theory/model}

\title{In Defence of Visual Analytics Systems: Replies to Critics}

\author{
Aoyu Wu, Dazhen Deng, Furui Cheng, Yingcai Wu, Shixia Liu, and Huamin Qu
}
\authorfooter{
\item
 A. Wu,  F. Cheng, and H. Qu are with the Hong Kong University of Science and Technology. E-mail: \{awuac, fchengaa, huamin\}@connect.ust.hk. 
\item
 D. Deng and Y. Wu are with the State Key
Lab of CAD\&CG, Zhejiang University. E-mail: \{ycwu, dengdazhen\}@zju.edu.cn. 
\item 
 S. Liu is with School of Software, Tsinghua University, China. E-mail: shixia@tsinghua.edu.cn.
\item The first two authors contributed equally to this work.
}

\shortauthortitle{Biv \MakeLowercase{\textit{et al.}}: Global Illumination for Fun and Profit}

\abstract{
The last decade has witnessed many visual analytics (VA) systems that make successful applications to wide-ranging domains like urban analytics and explainable AI.
However, their research rigor and contributions have been extensively challenged within the visualization community.
We come in defence of VA systems by contributing two interview studies for gathering critics and responses to those criticisms.
First, we interview 24 researchers to collect criticisms the review comments on their VA work.
Through an iterative coding and refinement process,
the interview feedback is summarized into a list of 36 common criticisms.
Second, we interview 17 researchers to validate our list and collect their responses,
thereby discussing implications for defending and improving the scientific values and rigor of VA systems.
We highlight that the presented knowledge is deep, extensive, but also imperfect, provocative, and controversial,
and thus recommend reading with an inclusive and critical eye.
We hope our work can provide thoughts and foundations for conducting VA research and spark discussions to promote the research field forward more rigorously and vibrantly.
} 


\CCScatlist{ 
 \CCScat{K.6.1}{Management of Computing and Information Systems}%
{Project and People Management}{Life Cycle};
 \CCScat{K.7.m}{The Computing Profession}{Miscellaneous}{Ethics}
}

\clearpage

\vgtcinsertpkg

\begin{document}
\firstsection{Introduction}
\maketitle
\begin{figure}[!t]
	\centering
	\includegraphics[width=1\linewidth]{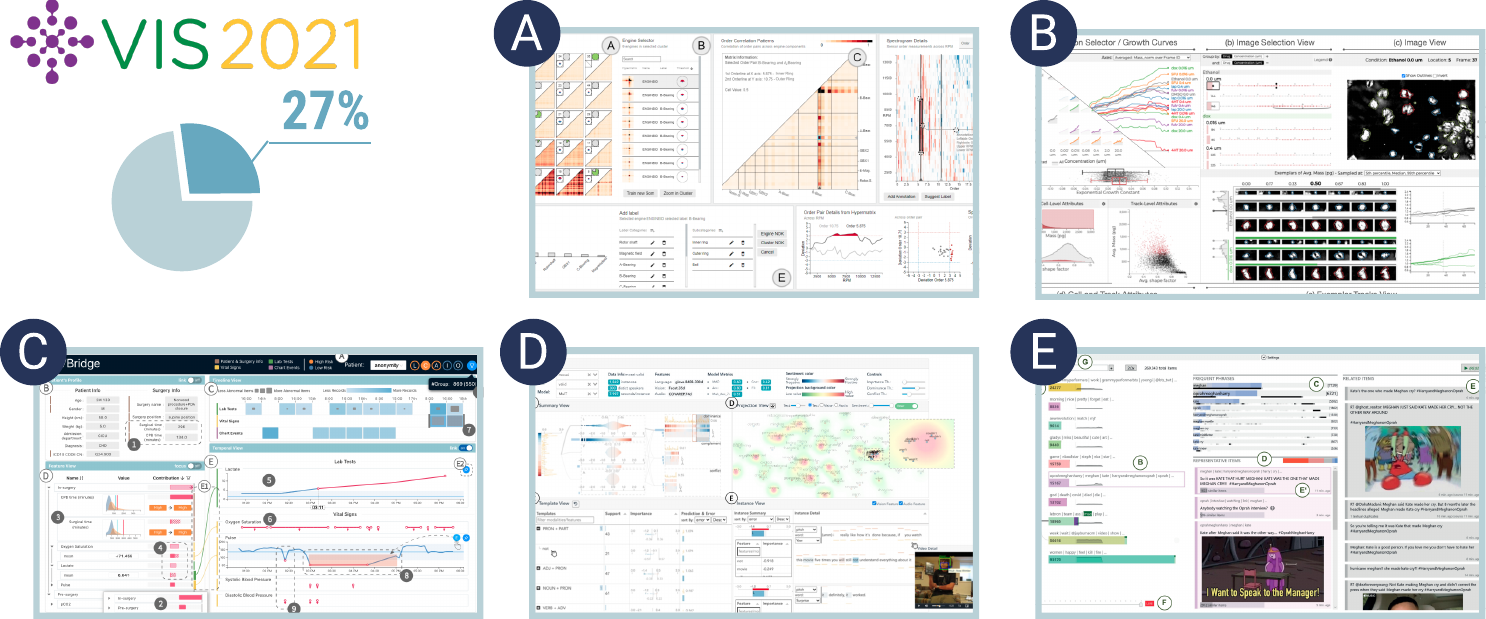}
	\caption{Research on VA systems accounts for 27\% (30/111) of the full paper published in IEEE Visualization Conference (VIS) 2021. Award-winning works include (A) IRVINE~\cite{eirich2021irvine}, (B) Loon~\cite{lange2021loon}, (C)~VBridge~\cite{cheng2021vbridge}, (D)~M2Lens~\cite{wang2021m2lens}, and (E) the system by Knittel~\ea~\cite{knittel2021real}.}
	\label{fig:vasys}
	\vspace{-15px}
\end{figure}

The last decades have witnessed extensive and ever-growing research interests on visual analytics.
By combining automated data analysis techniques with interactive visualizations,
visual analytics facilitates an effective understanding, reasoning, and decision-making on large and complex datasets~\cite{keim2008visual}.
Since the IEEE Conference on Visual Analytics Science and Technology (VAST) was founded in 2006,
researchers have contributed many systems for solving complex problems in wide-ranging applications,
to which we refer as \textbf{VA systems}.
Many systems are outcomes of problem-driven research,
whose values have been demonstrated through successful applications in high-impact domains, such as social media~\cite{wu2014opinionflow},
sports~\cite{wu2018forvizor,andrienko2019constructing},
urban analytics~\cite{ferreira2013visual,liu2016smartadp}, and explainable AI~\cite{liu2016towards,rauber2016visualizing}.
Research on VA systems have also become an important and impactful research field in visualization (\autoref{fig:vasys}).

Despite the success of those application-oriented VA systems,
their research contributions and rigor have been extensively challenged, discussed, and debated~\cite{meyer2019criteria,sedlmair2012design,weber2017apply,chen2019ontological}.
Underlying those debates is the tension between the impetus to create \textit{specific} software artifacts and the drive of academic research to produce \textit{general} knowledge~\cite{parsons2021understanding}.
This tension raises the frequently asked question of ``what our visualization community can learn from the VA system beyond solving the domain-specific problem.''
Furthermore,
the design and validation methodology of VA systems is human-centered and thus \textit{qualitative} and \textit{subjective} in nature.
This nature might contrast with the tropism in science and computer science that embraces \textit{quantification} and \textit{objectivity}~\cite{meyer2019criteria,isenberg2013systematic}.

The above discrepancies point out some practical problems in the research field.
For contributors,
their process of planning, developing, validating, and reporting VA systems is prone to diverse mistakes or pitfalls at different stages~\cite{sedlmair2012design}.
The process is challenging because conducting research on VA systems requires a wide range of skills, such as leveraging HCI approaches to understand target users, 
implementing automated algorithms,
and designing effective visual designs~\cite{thomas2009challenges}.
For reviewers, assessing the quality of VA system research requires judging and weighing the above aspects.
This assessing process tends to be subjective due to the lack of a shared ground among reviewers regarding objective criteria for evaluating VA systems~\cite{weber2017apply}.

We, therefore, seek to identify a common space of critics for VA systems and corresponding replies to critics.
We note that existing instructions on assessing the broad scope of visualization research are not specific to VA systems~\cite{lee2019beyond}.
Besides, the current reflections and discussions on the vigor of visualization design studies (\eg~\cite{meyer2019criteria,sedlmair2012design,meyer2018reflection}) are mainly constructed from authors' engagement and experiences.
While those discussions are undoubtedly critical and insightful,
we argue that surveying a broader scope of VA researchers will lead to a more diverse, representative, and convincing understanding.

We conduct two interview studies to reflect on VA systems in terms of their common criticisms and corresponding replies.
As shown in \autoref{fig:method}, we start by interviewing \np~researchers in charge of \nva~VA systems by asking them,
``what are the criticisms you have received during peer-reviewing?''
We obtain \ncrit~instances of criticisms,
which are further classified into \nc~common types through iterative categorization.
As those criticisms are diverse, deep, and challenging,
we conduct another interview study with \npsecond~researchers to validate our classification and gather their replies, \eg~how to mitigate and respond to those criticisms?
\begin{rev}
Drawing upon replies to low-level criticisms, we discuss implications for a high-level question - how to conduct research to defend and improve the research values and rigor of VA systems?
\end{rev}

We position our work as a preliminary probe into assessment criteria for VA systems.
We expect our work to benefit both VA contributors and reviewers by offering an evidence-based reference for assessing VA systems.
We, however, emphasize that criticisms are intrinsically biased, changing, controversial, and fallible, so is our result.
It is not our intention to advocate a golden template, but to construct a preliminary and debatable set of criteria for assessing VA systems.
We hope our work can provide foundations and spark discussions to make the research field more rigorous and vibrant.
We make our interview data and other supplemental material available at \href{https://re-vast.github.io}{\color{blue}\mintinline{html}{re-vast.github.io}}.





\section{Related Work}
We focus on research that primarily contributes a VA system for an application.
Thus, we review existing literature regarding theoretical advances in visualization application and design study, visual analytics,
and empirical methods for understanding the field of visualization.

\subsection{Visualization Application and Design Study}
Sedlmair~\ea~\cite{sedlmair2012design} formally introduced a visualization design study as ``a project in which visualization researchers analyze a specific real-world problem faced by domain experts, design a visualization system that supports solving this problem, validate the design, and reflect about lessons learned in order to refine visualization design guidelines''.
Following this definition,
they further proposed a nine-stage methodology framework for visualization design studies.
Their well-cited framework has become a common method of developing visualization systems for solving a domain-specific application and inspired many alternative design methodologies such as design activity framework~\cite{mckenna2014design}, design by immersion~\cite{hall2019design}, and design study lite~\cite{syeda2020design}.

However, concerns and debates have been raised about the research contributions and rigor of design studies and applications.
Sedlmair~\cite{sedlmair2016design} characterized 7 types of research contributions resulting from design studies.
Meyer and Dykes~\cite{meyer2018reflection} questioned how a specific design study might generalize to and benefit other visualization contexts,
advocating the need for developing standards to reflect on applied design studies to generate general knowledge.
Weber~\ea~\cite{weber2017apply} argued for the benefits and contributions of visualization application papers,
but called actions to develop criteria for how application papers can make clear and accessible contributions.

In response to those debates, Meyer and Dykes~\cite{meyer2019criteria} developed a set of six criteria for rigor in design studies: informed, reflexive, abundant, plausible, resonant, and transparent.
They took a deductive, top-down perspective by drawing conclusions from established criteria and principles in social science,
resulting in a set of criteria that is high-level.
In response, we adopt an inductive, bottom-up approach by observing criticisms in peer-reviewing to identify generality.
Our resulting set of criteria is thus low-level and specific.

\subsection{Visual Analytics}
A VA system is a software artifact applying visual analytics techniques.
It is common to decompose a VA system into two components,
namely data processing (mining) and interactive visualizations~\cite{keim2008visual},
while the latter is often further unfolded into visualization and interactions~\cite{chen2019ontological}.
VA systems extend information visualization systems by highlighting the usage of advanced data analysis algorithms to accomplish analysis tasks~\cite{keim2008visual}.
Thus, VA systems are considered to be complex~\cite{robertson2009scale,chen2019ontological}.

This complexity poses unique challenges in evaluating VA systems.
Scholtz~\cite{scholtz2011developing} analyzed reviews for entries to the 2009 VAST Symposium Challenge to develop an initial set of guidelines for evaluating VA systems,
which were further expanded into metrics including accuracy, the analytics process, the visualizations, and interactions~\cite{scholtz2014evaluation}.
Chen~\ea~\cite{chen2019ontological} proposed an ontological framework to evaluate VA systems by analyzing their symptoms, causes, remedies, and side effects.
In addition to understanding metrics, multiple studies aimed to survey evaluation methods.
Drawing upon surveys about evaluation methods in information visualization~\cite{lam2011empirical} and the whole visualization field~\cite{isenberg2013systematic},
Khayat~\ea~\cite{khayat2019validity} summarized seven common evaluation methods for VA systems and discussed their validity, generalizability, and feasibility.

We extend those discussions on evaluating VA systems to the assessment of research on VA systems.
Assessing the research values and rigor of VA systems requires not only judging the resulting VA systems,
but also the overall scientific process of planning, designing, validating, and reporting VA systems.
We note that evaluating VA systems remains a challenging and controversial issue,
and present our findings of common criticisms on the evaluation.

\begin{figure}[!t]
	\centering
	\includegraphics[width=1\linewidth]{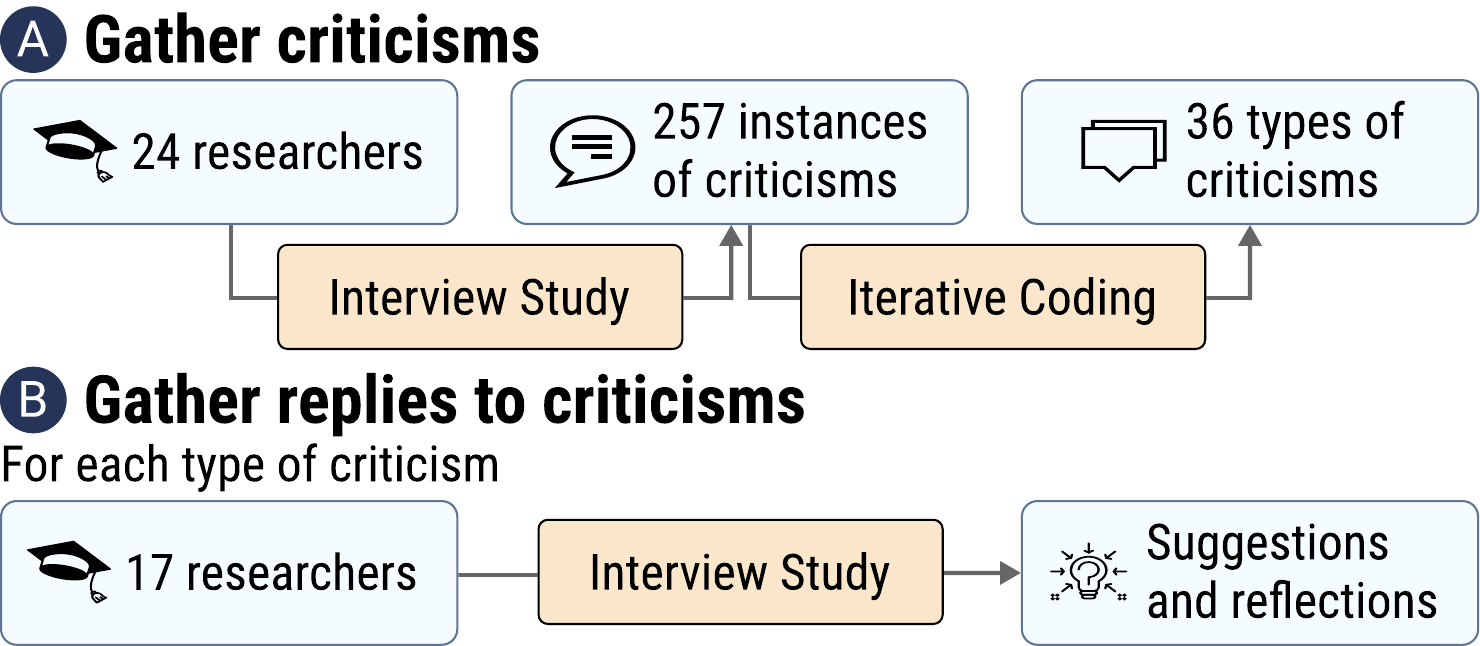}
	\caption{Our study consists of two phrases including (A) identifying common criticisms and (B) gathering replies to each type of criticism.}
	\label{fig:method}
	\vspace{-15px}
\end{figure}

\subsection{Understanding the Field of Visualization}
It has been common to survey visualization literature to understand the field of visualization.
For example, Isenberg~\ea~\cite{isenberg2016visualization} analyzes the keywords in visualization papers to identify research topics and trends,
which is later integrated into a meta collection of visualization literature~\cite{isenberg2016vispubdata}.
Lee~\ea~\cite{lee2019broadening} summarized 25 common contributions of visualization research.
Besides,
researchers have contributed many surveys about specific issues of visualizations such as evaluation methods~\cite{lam2011empirical,isenberg2013systematic},
design spaces (\eg~visualization tasks\cite{schulz2013design}),
and trending topics (\eg~AI4VIS~\cite{wu2021ai4vis} and ML4VIS~\cite{wang2020survey}).

Another line of research focuses on surveying visualization researchers to gather insights on trending topics such as immersive analytics~\cite{ens2021grand} and big data visual analytics~\cite{andrienko2020big}.
Furthermore, some visualization approaches have been developed to understand the research profiles~\cite{latif2018vis} and career paths~\cite{wang2021seek} of visualization researchers.

Our work contributes a new practice of surveying visualization researchers in the context of peer reviews.
We leverage the dual roles of researchers as both contributors (authors) and reviewers to ask the probably tasteless and consequentialism-oriented questions ``what are the criticisms you received from peer reviews'' and ``how to react to those criticisms.''
This practice allows us to gather a rich and diverse corpus of professional opinions,
that is deep, extensive, but also nuanced and contradictory.
We hope that our practice could stimulate dialogue and debate around peer reviews that are vital to our research community.



\section{Methodology}
Our work consists of two interview studies with researchers on VA systems.
As shown in~\autoref{fig:method},
we first interviewed researchers to collect instances of criticisms that they received in their experience of contributing VA systems,
which were further classified into \nc~categories through iterative open coding.
In the second study,
we interviewed \npsecond~researchers to understand their replies to those criticisms.

\subsection{Study 1: Gathering Criticisms}
We conducted structured interviews with researchers on sample criticisms of VA systems. The word ``researcher'' refers to researchers that have both contributed and reviewed at least one paper 
that falls into our scope (\ie work where the primary contribution is a VA system) 
in the IEEE Visualization Conference (VIS) or IEEE Transactions on Visualization and Computer Graphics (TVCG).

\begin{figure}[!t]
	\centering
	\includegraphics[width=1\linewidth]{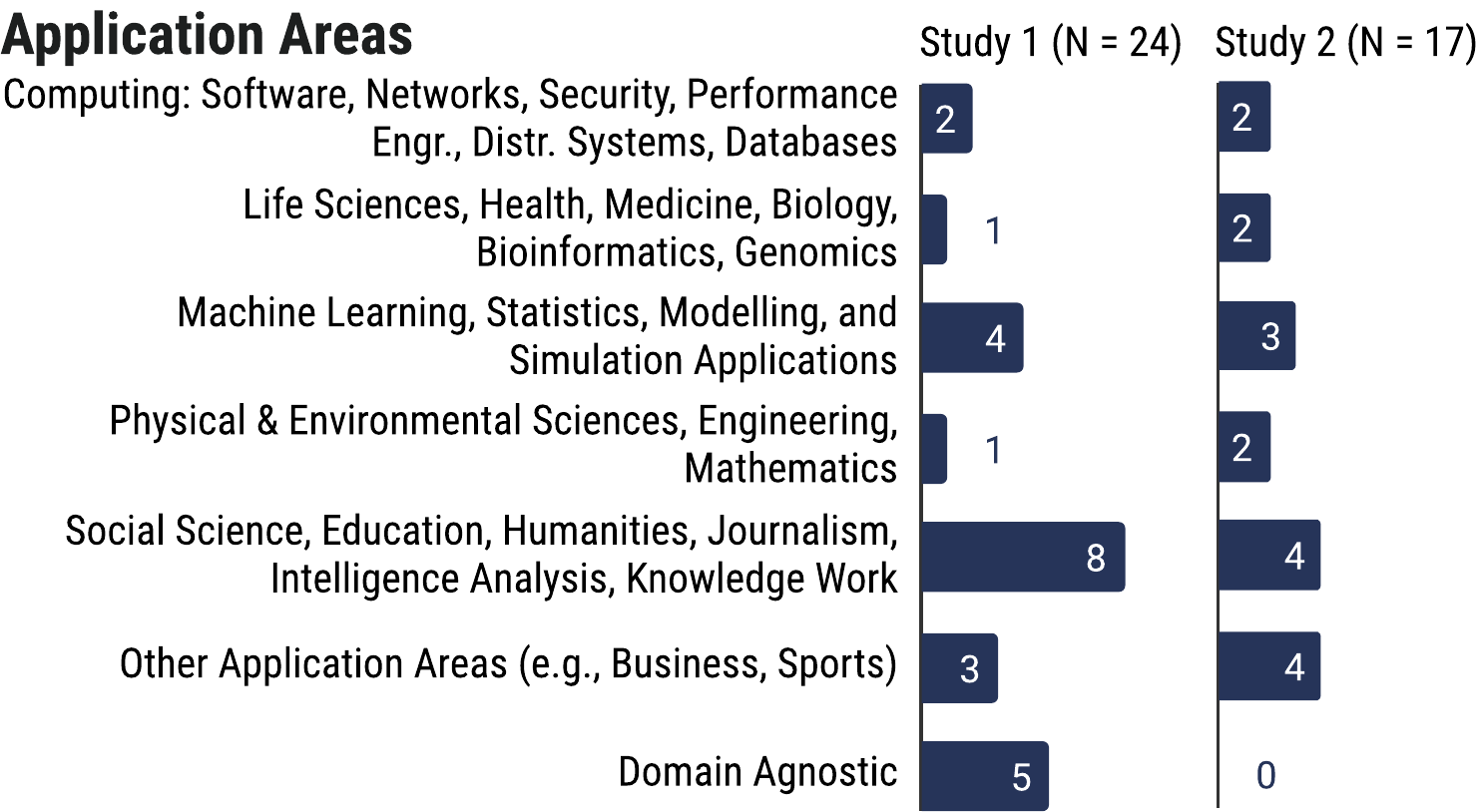}
	\caption{Application domains of participants in interview studies.}
	\label{fig:participants}
	\vspace{-15px}
\end{figure}

\textbf{Participant.} We interviewed \np~participants (including 7 Ph.D. students, 7 postdoc researchers, 7 research scientists, and 3 professors) who were reported to have authored a total number of \nva~research papers as the first author.
We recruited them from personal and professional connections as well as emails.
\begin{rev}We reported on the participants' application domains as categorised in VIS Paper Submission Keywords~\cite{viskeywords} and types of work.
As shown in~\autoref{fig:participants}, they covered a range of areas.\end{rev}

\textbf{Interview.}
We conducted structured interviews with individual participants,
asking them about their experience of receiving criticisms for their VA systems.
We were aware of concerns over the copyrights and anonymity of peer reviews, and thus purposeful requested participants not to quote the reviews but instead reported on aggregated understandings based on all their research instances.

\textbf{Analysis.}
We iteratively open-coded interview data and applied an informative structure to organize and report the findings, \ie grouping the criticisms by their corresponding components or sections in the manuscript (\autoref{fig:sections}).
The major consideration of choosing this grouping scheme is the intuitiveness and the strong connections between writing components and design stages~\cite{sedlmair2012design}. 
To overcome the phraseology inconsistencies of writing components in different interviews (e.g., ``design requirements'' and ``design goals''), we summarized the common writing components from existing VA research papers. 

Three authors labeled and categorised those criticisms iteratively, and discussed conflicts until reaching a consensus.
This iterative process yielded several local adjustments about splitting, merging, and removing items in the list.
For example, ``not working with real experts'' was split from ``unclear evaluation methods'' and merged with ``not using real data" into "lacking realism''.
We provide logs of changes in the supplemental material to encourage future discussions.

\subsection{Study 2: Gathering Replies}
With the above list of criticisms,
our next goal was to validate the list and provide feedback in response to each criticism.
Different from previous work that offered suggestions on writing or reviewing visualization papers based on authors' experience (\eg~by Munzer~\cite{munzner2008process}, Stasko~\cite{Stasko16}, and~Elmqvist~\cite{Elmqvist16}),
we sought to be more objective and evidence-based by interviewing active researchers.

\textbf{Participant.}
We adopted a similar method to recruit participants as in the previous study.
We recruited 17 participants (5 females), including 6 senior Ph.D. students, 3 postdoc researchers, 4 research scientists, and 4 professors.
They had published a sufficient number of papers in IEEE VIS/TVCG (Mean: 6.6, SD: 3.8).
We additionally collected their career ages~\cite{tovanich2021gender},
which were the number of years since the author published their first paper in our scope (Mean: 5.3, SD: 1.9).

\textbf{Interview.}
We conducted structured interviews with individual researchers, each lasting for 1 to 2 hours.
\begin{rev}
We asked participants pre-defined questions in a list,
since it allowed us to gather focused feedback on how to address criticisms and improve scientific rigor.
We also encouraged them to share any opinions to avoid missing thoughts and explicitly asked them whether they had additional comments that were not covered in our list.  
\end{rev}
Our interview questions were designed surrounding two research questions ($\mathcal{Q}$1-2) and were classified into two types: one-off (O) and repeated for each criticism (R).

$\mathcal{Q}1$ - Is our list mutually exclusive and collectively exhaustive?
\begin{compactitem}
    \item Did you encounter this criticism when serving as VA system contributors or reviewers, respectively? (R)
    \item Do you think this criticism can be merged with others? (R)
    \item Did you encounter any other criticisms that are not listed? (O)
\end{compactitem}

$\mathcal{Q}2$ - How can VA researchers respond to those criticisms?
\begin{compactitem}
    \item How important is it? What are the recurring pain points? How to avoid or address it? (R)
    \item \begin{rev}On a 7-Likert scale, is the criticism not at all (1) or extremely specific (7) to VA systems (R)?\end{rev}
\end{compactitem}




\section{Criticisms and Replies}
\label{sec:pitfall}
\begin{figure}[!t]
	\centering
	\includegraphics[width=.9\linewidth]{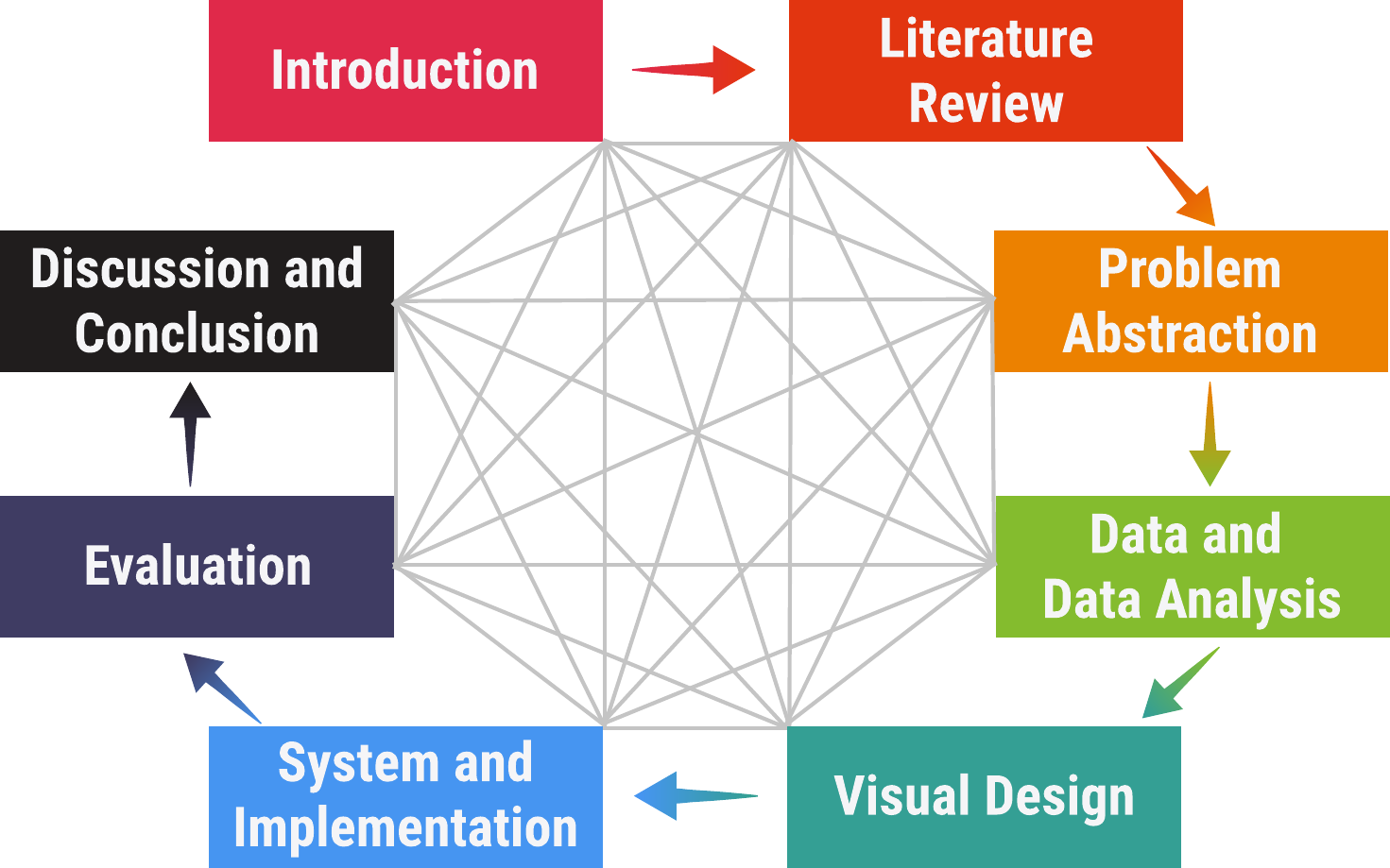}
	\caption{Eight components of VA systems manuscripts. While the overall structure is sequential, each component is linked with others.}
	\label{fig:sections}
	\vspace{-15px}
\end{figure}

In this section, we first describe the one that relates to all the eight components in~\autoref{fig:sections}, then discuss others following the order.
We use representative quotes from interviewees in the second interview study (denoted P1-17) throughout the section to support our claims.
\begin{rev}
The background color encodes corresponding components as in~\autoref{fig:sections}. 
\end{rev}


\crit{dGrey}{lGrey}{Lacking links of coherence and consistency}
As shown in~\autoref{fig:sections},
the eight components are ordered sequentially but interlinked.
Missing the links of coherence and consistency is the first issue.
There are diverse cases of inconsistency:
claims about novelties are not substantiated by comparisons with related work;
participants in the evaluation do not conform to proposed user groups in problem abstraction and task analysis.

Tackling this problem would require ``following a systematic approach to design, develop, and validate VA systems'' (P3) and ``logical writing'' (P6).
P4 advocated that researchers clearly illustrate the connections between domain problems, analysis tasks, VA designs, and evaluations. 
To this end, he suggested using ``layered graph models'' to guide and report research processes, where nodes in each layer represent different components.
Correspondingly, one can draw edges to denote relationships, \eg~a view accomplishes some tasks,
enabling graph-based analysis such as conducting coverage tests and finding isolated nodes.
We suggest future research to propose formalisms and evaluate them to progress the theorization of VA systems.

\subsection{Introduction}
\label{sec:pitfall:intro}
The introduction component results from careful consideration and organization of the overall research during iterative paper writing.
Thus, issues in this component are high-level and linked with other sections.
As such, they need to be considered throughout the whole research.

\crit{dRed}{lRed}{Unclear relevance to visual analytics}
The first step of reporting research results is to provide the background information that motivates the research.
In our context, it is important to articulate why this is a visual analytics problem.
A common motivation scenario is that ``automated methods do not solve the problem'' (P10),
bringing the need to ``integrate computational methods with humans through interactive visualizations'' (P6) to make decisions.
That said,
visual analytics problems are not about ``visualizations of computed results'' (P6).
Furthermore,
it could be a pitfall by simply stating that there is no prior VA system for the domain problem.
No previous work does not necessarily lead to the necessity of VA systems,
since it can also because the domain problem does not need to loop humans in the analysis process.
Thus,
it is essential to articulate the challenges of domain problems and the limitations of existing solutions.

\crit{dRed}{lRed}{Vague/over-claimed contributions}
The ending part of an introduction is often dedicated to a brief summary of the contributions.
This has been a common practice according to our survey,
\ie only one paper~\cite{tovanich2021miningvis} does not list the contribution.
A clear statement of contributions is particularly vital for VA systems research,
whose contribution types can be very diverse.
Besides,
the stated contributions need to accord with the targeted problems.
Because VA systems are often designed for specific problems,
one should not ``over-claim the scope of the application'' (P12).


\begin{table*}[]
\small
\caption{Classifications and statistics of contribution types found in 30 surveyed papers in IEEE VIS 2021.}
\label{table:contributions}
\centering
\begin{tabular}{l|l|l}
\toprule
\textbf{Contribution Type} & \textbf{Count} & \textbf{Example (Quoted from the Text)}  \\ \hline
Evaluation & 27 & The evaluation of IRVINE together with six automotive engineers~\cite{eirich2021irvine}. \\ 
VA system/tool/prototype & 23 & ThreadStates, an interactive web-based tool for state-based visual analytics of disease progression~\cite{wang2021ThreadStates}. \\ 
VA design/workflow/framework & 17 & We introduce a risk-aware framework, namely, VideoModerator, to facilitate the efficient moderation of e-commerce videos~\cite{tang2021videomoderator}. \\
Visual representation & 10 & New visualization methods were developed to assist users to further understand critical factor data~\cite{maher2021ffective}. \\ 
Design study (Problem abstraction) & 9 & We characterize the problem domain of visual analytics of time-varying effects of multiple factors on academic career success~\cite{wang2021seek}. \\ 
Data mining algorithm & 8 & We propose a dynamic clustering algorithm to enable the efficient clustering of fast-paced incoming streaming data~\cite{knittel2021real}. \\
Open-sourced implementation & 3 & An open-sourced, web-based implementation ... (omitted)~\cite{park2021neurocartography}. \\ 
Reflection & 2 & A start-to-end description of the lessons learned from this successful, multi-site remote collaboration~\cite{floricel2021thalis}. \\ 
New domain and problem & 2 & The first visual analytics framework for diagnosing and improving deep semantic segmentation models ... (omitted)~\cite{he2021can}.\\
Data model & 2 & A technique to model users' analytic behavior from interactions with the data~\cite{narechania2021lumos}. \\ 
Dataset & 1 & A dataset of scraped metadata from 59, 232 academic articles~\cite{narechania2021vitality}. \\ 
\bottomrule
\end{tabular}
\begin{tablenotes}
  \small
  \item Remark: One research paper~\cite{tovanich2021miningvis} does not explicitly claim the contribution and is excluded from this analysis.
\end{tablenotes}
\vspace{-15px}
\end{table*}


\crit{dRed}{lRed}{Unclear contributions to the VA community}
We observe concerns regarding how the contributions are relevant to the VA community,
\eg~``what our community can learn from this research beyond solving the domain-specific problem''?
This criticism sparked extensive discussions during our interviews,
promoting us to survey what contributions had been claimed (see the supplemental material for details).
As shown in~\autoref{table:contributions},
the most frequent claimed ones are the evaluation (27),
followed by VA systems/tools/prototypes (23),
and VA design/flow/framework (17).
Those contributions are intrinsically specific to the domain problem.
Our interviewees generally agreed that solving the domain problem is the primary contribution,
but they also expressed a clear need for ``generating new knowledge for visualization researchers'' (P6).

What might generalize to other problems are novel visual representation (10) and data mining algorithms (8) capable of certain data types and analytical tasks.
Characterizing the domain problem (9) might also inspire future research in the domain,
especially when applying to application domains that were not previously reported in the visualization literature (2),
\ie~``It is a significant contribution to firstly apply VA systems to a new application domain'' (P2,4).

There are other opportunities for providing general knowledge for visualization researchers,
including open-sourced tools (3), reflection (2), data model (1), and dataset (1).
Our interviewees also outlined other possibilities,
including generalized ``advice for the visualization design'' (P6) and
demonstration of the generalizability to other datasets or application domains (P12,14).
We recommend our readers think creatively and broadly about other possible valuable contributions.

\crit{dRed}{lRed}{Unclear novelty of VA systems}
The novelty of the VA system plays an important role,
\ie~``a grab-bag of existing techniques is hard to be appreciated by visualization researchers'' (P1).
The novelty might not necessarily be novel data mining techniques or visual designs,
but instead ``lies in the workflow, in particular how the VA workflow is integrated into the domain workflow'' (P5). 
Similarly, P16 commented, ``I usually consider the paper has sufficient novelties if it has novelty in any part of the VA pipeline, e.g., data processing, user interaction, visual design, user study experiment design and etc.''
To demonstrate the novelty,
it is often necessary to ``qualitatively compare with existing VA systems'' (P12).
We will discuss quantitative comparisons in~\autoref{pitfall:eva}.


\subsection{Related Work}
Composing the related work section prompts authors to identify relevant research topics,
survey and summarize publications to shed light on gaps and articulate novelties.

\crit{dRL}{lRL}{Missing related work}
Missing related work is a common criticism.
Although it is often considered a minor issue, it requires a thorough understanding of the core contributions and relevant research topics.
This can be difficult since a VA system concerns a wide range of topics, such as the domain problem, automated and interactive VA solutions to the domain problem, and visualizations for the abstract data (\eg~text visualization) and analytical tasks (\eg~visual cluster analysis).
P15 suggested building a database to organize existing VA systems ``from different perspectives such as application domains and data types'' to help communicate the differences and commonalities.

\crit{dRL}{lRL}{Inadequate discussion about related work}
Writing the literature review is not an enumeration of relevant publications,
but instead a critical organized account of the current state of research and knowledge.
The results need to be synthesized into a summary of what is known and unknown in the research field,
and how the new VA system advances the field.
Lacking the depth of critical discussions would lead to an issue that fails to inform readers of the current research frontier.
Besides, it is crucial to provide ``a qualitative comparison with closely related systems'' (P10).

\subsection{Problem Abstraction}
This section formally signifies the entrance to the design study,
\ie working with domain experts to understand the domain problem, collect user requirements, and abstract the data and analytical tasks to inform the design of VA systems.

\crit{dPro}{lPro}{Unclear domain experts or target users}
An immediate question arises regarding the profiling of experts and users.
We differentiate end-users from domain experts. 
For example, when designing a learning tool for students, teachers play the role of domain experts, helping researchers characterize domain problems.
It should be made clear 
whether they are real users or fictional personas,
what are their working fields and domains,
and what are their required knowledge to interact with the system.
Regarding this criticism,
P4 suggested that it could help to clarify ``their knowledge about visualization'' since VA systems may need visualization expertise.

\crit{dPro}{lPro}{Unclear methodology and methods for problem abstraction}
In addition to domain experts,
it is important to adopt a systematic method for problem abstraction and provide sufficient details.
P5 emphasized that this is a common problem, 
saying ``many submissions do not detail this step. Understanding the domain problem is difficult and requires iterative collaboration with experts.''

According to Sedlmair~\ea~\cite{sedlmair2012design}, a methodology is like a recipe describing ``strategy, plan of action, process, or design lying behind the choice and use of particular methods`` and methods are like ingredients.
The common methodology in our surveyed papers includes design study (\eg~\cite{eirich2021irvine,cheng2021vbridge,tang2021visualization}),
user-centered design and its extension (\eg~\cite{pu2021matexplorer,wang2021ThreadStates,floricel2021thalis}),
and the nested model of visualization design~\cite{wang2021seek}.
Methods for understanding the domain problems are primarily interviews (discussions and meetings) and literature review.
Other surveyed methods include contextualized design (\eg~one author embedded himself in the domain experts' research group for one year~\cite{lange2021loon}),
workshops~\cite{lange2021loon}, formative and pilot studies (\eg~\cite{cheng2021vbridge,tang2021videomoderator}).
We note that those methods are not exhaustive and encourage readers to explore other approaches such as ``fly-on-the-wall'' (P6).

Furthermore, it is important to provide a structured discussion about the domain problems.
For instance,
what are the current workflows and practices of domain problems?
What are the challenges encountered by domain experts?
Providing such information could facilitate understandings by readers who are not familiar with the specific domain,
\ie~``why do experts have those problems and what kind of tools can help them'' (P10,12).
Besides, as the design study is often iterative,
domain problems can be revised throughout the design process.


\crit{dPro}{lPro}{Insufficient abstraction from domain to VA problems}
With the derived domain problems and requirements,
it is essential to perform data abstraction and task abstraction to inform the design of the VA system.
While domain problems are often expressed in domain-specific language,
data and visualization tasks need to be described in visualization language to communicate the relevance to visualization researchers.
Such abstraction could make the visual design potentially applicable to other domain problems.

Many interviewees stressed the challenges of abstraction for many reasons:
different domain problems have varying levels of granularity which might be difficult to abstract (P1);
domain problems are even not well-defined (P4,8);
it is hard to differentiate between domain-specific and visualization-specific terms (P2,6);
and different papers describe similar visualization problems in different languages (P11,15).
Their comments call for actions to organize exiting practice of problem abstraction to clarify the terminology, identify common patterns and topology, and establish common vocabularies to guide future research.



\subsection{Data and Data Analysis}
Developing VA systems typically starts with processing the raw data and developing models (\ie data analysis or mining models) for automated analysis~\cite{keim2008visual}.
We collectively refer to them as data and data analysis.
Data analysis is not standalone in VA systems but is tightly integrated with interactive visualization.
For this reason, we find it is difficult to completely seclude criticisms in this component from those in visual design (\autoref{criticism:visual}) and system (\autoref{criticism:system}).
In addition, we observe a large number of technical problems that are closely related and specific to the domain problems.
Therefore, we exclude those specific technical problems from our discussions.

\crit{dData}{lData}{Unclear definitions and explanations of data}
This section usually starts with explaining the data and data-related issues such as definitions, metrics, and features.
This is particularly important when the datasets are not public and ``the systems are not designed for common types of data'' (P17).
In addition to plain text description, a useful approach is to ``provide an example of the dataset in a table or figure'' (P2).
It is suggested to describe the characteristics of the abstract data,
such as the data type (\eg relational or graph) and the number of dimensions.
Such information might help readers understand whether the VA system could generalize to other datasets.

\crit{dData}{lData}{Missing technical details}
Following the data,
one should describe technical details for the subsequent data analysis process (\eg~data processing, mining, and learning).
The goal is to enable readers to comprehend and replicate the methods by following the description.
However,
we find this criticism to be common in our surveyed sample.
Part of the reason is the ``curse of expertise''~\cite{fisher2016curse} that experts tend to overestimate their ability to explain their areas of formal expertise.
Another reason might be the ``limited page lengths that make it hard to provide all details'' (P4).
Alternative approaches are to describe details in the supplemental material, to provide source codes with documentation, or to provide a figure with a concrete example that illustrates the algorithm step by step (P15).

\crit{dData}{lData}{Novel data processing algorithms}
As shown in~\autoref{table:contributions},
contributions of VA systems can include novel data processing algorithms.
However,
we observe diverse opinions about such contributions.
On the one hand, some interviewees considered them to be an ``addition'' that were not necessarily well-targeted for the visualization community and difficult to assess (P11,14),
\ie~``if they are standalone contributions, they should be split into another paper to other venues'' (P6).
On the other hand,
P5 appreciated such contributions,
saying ``analysis methods are an important part of visual analytics.''
A helpful opinion to reconcile this conflict may be ``it depends on whether the algorithm is closely integrated into the VA systems, for example, to support real-time interaction for streaming data'' (P6).
We report on those controversies in an attempt to spark constructive discussions.
From our perspective,
we echo the calls for ``broadening intellectual diversity in visualization research papers''~\cite{lee2019broadening} and argue that the visualization community should stay inclusive and open to computational analytics techniques.
However, we note that such contributions might be challenging to assess, thus advocating involving data analysis professionals during peer reviews.


\crit{dData}{lData}{Unclear choices of algorithms/models}
It is certainly legitimate to adopt existing methods for data analysis.
However, 
the choices should be justified,
\ie~``the WHY question is more important than the HOW question for researchers'' (P13).
The purpose is to convince that the selected methods are solid and adequately applied to solve the particular problem,
\ie~``the decision should not be arbitrary'' (P4).
Suggestions include ``abstracting the problem, for example, clustering or frequent pattern mining'' (P7), ``clarifying the input and output'' (P6), followed by explaining candidates and decisions.



\subsection{Visual Design}
\label{criticism:visual}
The core component of VA systems is the interactive visualization that integrates automated data analysis with human analysis~\cite{keim2008visual}.
Because both computational data analysis and interactive visualization are techniques,
we find criticisms in visual design to be similar to that in data and data processing,
\ie missing details, unclear novelties or design choices.
However, as visual designs are often often the core interest to the visualization community,
we also find several specific criticisms.

\crit{dVis}{lVis}{Missing details of visual designs}
A clear description of the visual designs makes the complex design, sometimes with multiple views, more comprehensible to readers.
To this end, a brief overview of the whole visualization is often required.
For each view,
the description includes the input data, data transformation, and visual mappings.
This description can be followed by explaining the interactions and coordination among views.

Nevertheless,
it remains an open challenge to explain visual encodings of common data visualizations effectively~\cite{yang2021explaining}.
It is, therefore, far more challenging to explain visual encodings of VA systems that are considerably more complicated.
Multiple interviewees emphasized that the figure should provide legends (P2,4,7,8), and the demonstration videos should explain the encodings (P12,13).
We encourage empirical research to investigate guidelines for explaining complex visualization design and making them more understandable (\eg~\cite{shu2020makes}).

\crit{dVis}{lVis}{Unclear novelty of visual designs}
Although novel designs are not a necessary product from design studies,
our interviewees confirmed that the criticism of ``lacking novelty of visual designs'' was still commonly raised in peer reviews.
As a reaction,
some interviewees argued for de-emphasizing novel designs.
P13 said, ``in the early years, I used to raise this concern. But now I think it is not critical.''
P17 further challenged, ``the goal of VA system is to solve the problem. Novel visualizations might not work in real-world scenarios.''
P4 challenged the death of novel visualizations, commenting ``our field has explored the design space extensively. There does not leave much space for novel designs.''
P10 further raised that ``the judgment (of novelties) can be subjective. There lacks a database to assure the novelty.''
Those opinions underscore the importance of continuing discussions on the assessment criteria for VA systems and developing a database to benchmark the state-of-the-art systems.

\crit{dVis}{lVis}{Unclear rationale and justifications of visual designs}
We find this criticism to be the most common one in our survey.
This finding is not surprising since visual design is of core interest to the visualization community,
\ie~``VA is a design problem. Without justifications, it is hard to convince'' (P12).
However, justifying visual designs requires solid argumentation,
spanning diverse issues such as whether the visual channels are effectively encoded,
whether the design is mostly approximate for the task,
whether designs are consistent among views,
and what are the potential pitfalls (\eg visual clutter).
These multiple-criteria decision-makings often lead to trade-offs that require careful consideration and strong justification.

Despite the importance as seen in peer reviews,
our interviewees suggested a lack of theoretical research on how to justify visual designs and judge the rigor.
They had adopted and seen a variety of methods,
such as referring to visualization design guidelines and well-established principles,
providing evidence that the design was chosen and proven by experts,
and discussing alternative designs.
However,
they also expressed concerns that existing practices are mainly based on argumentation or anecdotal evidence.
Therefore, it is important to develop systematic methods to improve scientific rigor.
To that end, P15 said, ``I wonder whether we can derive general knowledge from existing justifications to find common design patterns''.

\crit{dVis}{lVis}{Problematic visual designs}
An extreme case of the above criticism is that the visual design is considered to be problematic.
While concerns on C-17 are mostly ``soft'' (\ie~designs require further justifications),
this criticism is ``hard'' (\ie~designs have to be revised and improved).
For example,
the design violates well-established guidelines without necessary explanations,
\eg~using rainbow color maps.
Such problems can be surfaced by consulting experience visualization researchers.

\crit{dVis}{lVis}{Missing discussions about alternative designs}
As discussed in C-17,
interviewees thought of discussing alternative designs as one method for justifying design choices.
The underlying problem is whether the chosen visual design is the most appropriate one among many possibilities.
Successful designs typically require starting with a broad consideration space of possible solutions and subsequently narrowing proposal space~\cite{sedlmair2012design}.
Thus, comparing with alternatives provides more evidence on the validity of design decisions~\cite{liu2019understanding},
\ie~``without this step, the design progress might be unsystematic and weak'' (P7).
However,
some interviewees commented that this design progress was often neglected,
expressing the demands for better tools for facilitating the exploration, comparison, and management of alternative designs.

\crit{dVis}{lVis}{Over-complicated visual designs}
We find a considerable amount of issues challenging that the visual design is over-complicated.
We observed mixed opinions from our interviewees.
On the one hand,
some argued that the complexity was unavoidable due to the complex domain problems (P1,2,4,10).
The study should be treated as a success if users' workflows are improved with the complex VA systems; even the solution looks not ``elegant''.
Thus, they linked this criticism to insufficient justifications (C-17).

On the other hand,
more interviewees stressed that complexity could pose threats to usability,
\ie VA systems may be too difficult for users to learn and use (P3,5,6,11-15).
They further imputed over-complexity to the pursuit of novel designs,
\eg~``novel designs are being more and more complex'' (P11).
P12 commented, ``our field has undergone the evolution from simple to complex designs and I think now it is the time to reverse - to reflect on existing designs and pursue simplification.''
Therefore,
it is promising to study systematic methods for simplifying VA systems,
\eg~by finding unnecessary components through coverage testing~\cite{yang2009survey}. 
This perspective also raises new questions about measuring the complexity of visualization and VA systems.



\subsection{System and Implementation}
\label{criticism:system}
We refer to the system as the compilation of the data, algorithms, and visual design.
In this section, we discuss system-level criticisms.

\crit{dSys}{lSys}{Lacking workflow overviews or system demonstration}
Because the system can be complicated,
readers often need an overview of the overall system and the workflow to understand how all components work together.
There are many methods, such as providing an illustrative figure and describing usage scenarios as a walk-through for workflow.
P17 said, ``without clear introductions to the connections between views, it is hard to understand how the whole system works.''
Besides,
a system demonstration in interactive software or videos can provide an overview of the system to readers.

\crit{dSys}{lSys}{Uncertainty/stability/sensibility}
We observe the emergence of issues regarding uncertainty, stability, and sensibility, especially to the choice of algorithm parameters.
Those issues can root in the chosen algorithms,
\eg~different initial conditions in analysis algorithms, such as t-SNE~\cite{van2008visualizing}, can result in a significant difference,
posing threats to the reliability of the VA workflow.

\begin{table*}[]
\small
\caption{Evaluation strategies and collected data in 30 surveyed paper in IEEE VIS 2021. Numbers in brackets indicate the counts and numbers in braces denote the mean and standard deviation of the number of participants. References are example instead of full instances.}
\label{table:eva}
\centering
\begin{tabular}{l|p{7.8cm}|p{6cm}}
\toprule
\textbf{Strategy} & \textbf{Definition} & \textbf{Collected Data} \\ \hline
\parbox[t]{2.7cm}{Observational studies (22) \\
 $ \{N: 4.7 \pm 3.5 \}$
} & By observing how real users work with the VA system in the field or in the laboratory, \eg~conducting case studies and expert interviews~\cite{cheng2021vbridge}. This strategy also includes longitudinal field studies~\cite{tovanich2021miningvis}. & \parbox[t]{7cm}{qualitative subjective opinion, \eg~interview (21), \\ analytic workflow and insights* (15), \\
quantitative subjective opinion, \eg~questionnaire (6), \\ quantitative objective data (3): logs (2), task performance (1) } \\ \hline
Usage scenarios (12) & By describing how the VA system could be used by hypothetical users~\cite{pu2021matexplorer}. & analytic workflow and insights* (12) \\ \hline
\parbox[t]{2.7cm}{Demonstration (5) \\
 $\{N: 5.6 \pm 4.4\} $
}  & By demonstrating the VA system to the users, 
\ie~users do not work with the VA system but might ``explore it freely''~\cite{song2021interactive}. &  \parbox[t]{7cm}{qualitative subjective opinion (interview) (5), \\ 
quantitative subjective opinion (questionnaire) (1)} \\ \hline
Model experiment (5) & By analyzing the model (\ie data processing and data mining algorithms) performances~\cite{knittel2021real}. & \parbox[t]{7cm}{algorithm performance, \eg~accuracy and running time (4), \\
insight quality (1)} \\ \hline 
\parbox[t]{2.7cm}{Experimental studies (3) \\
 $\{N: 18.0 \pm 8.5\} $
} & By conducting controlled experiments where users work with the VA system versus baseline systems~\cite{tang2021videomoderator}. & 
\parbox[t]{7cm}{qualitative subjective opinion, \eg~interview (2), \\ 
quantitative subjective opinion, \eg~questionnaire (2), \\ 
quantitative objective data (2): logs (1), task performance (1) } \\
\bottomrule
\end{tabular}
\begin{tablenotes}
  \small
  \item *: Analytic workflow and insights are considered to be qualitative and either subjective or objective~\cite{khayat2019validity}.
\end{tablenotes}
\vspace{-15px}
\end{table*}

\crit{dSys}{lSys}{Scalability}
Scalability is a commonly raised concern,
questioning how the system, including data processing algorithms and visualization,
scales to the increasing amount of data.
This concern is mainly because visual analytics is motivated by ``the rapidly increasing amount of data''~\cite{keim2008visual},
which is further promoted during the big data era~\cite{keim2013big}.
It is, therefore, often expected that VA systems can handle massive volumes of data.
However,
interviewees noted that it depended on the typical data size in the application domains.
Some commented that it was not yet a very common practice to conduct quantitative experiments for system scalability,
advocating the use of software testing strategies to surface the scalability problem (P5,9).

\crit{dSys}{lSys}{Generalizability}
Generalizability appears as a common concern,
and many interviewees considered it vital,
\ie~``design study papers usually are more tailored to a specific application. I generally would prefer systems that provide a good abstraction of the data to make it generalizable to other applications.'' (P16).
On the one hand,
interviewees agreed that many VA systems were designed for a specific domain problem,
and ``generalizability is not their goals'' (P5).
On the other hand,
interviewees found poor generalizability a universal concern of VA systems.
Thus,
it is important to study how to promote the generalizability of VA systems so that ``our field can continue building up and accumulating knowledge'' (P6).
Many participants acknowledge that this question was challenging and warrants future research.



\crit{dSys}{lSys}{Usability}
Usability assesses how easy the visual interface is to use.
Criticisms on usability are tightly interwoven with those on over-complicated visual designs (C-20),
doubting that the design is complex enough to raise questions on learning curves and real-world usability.
However,
P6 noted that ``few VA systems pay enough attention to this problem and evaluate the usability in the field through longitudinal studies'',
which is worthy of more research attention and effort.


\subsection{Evaluation}
\label{pitfall:eva}
Evaluating visualization systems has been traditionally difficult~\cite{isenberg2013systematic},
and it is arguably even harder to evaluate complex VA systems.
To guide our discussions,
we start with surveying existing evaluation methods for VA systems (see the supplemental material for details).
The survey serves as an extension of existing literature from InfoVis~\cite{lam2011empirical}, VAST~\cite{khayat2019validity}, and VIS~\cite{isenberg2013systematic}, more focusing on evaluations of VA systems. 

As shown in~\autoref{table:eva},
observational studies in the field or in the laboratory (\eg~case studies) are the most common evaluation strategy.
In observational studies,
a variety of data can be collected and analyzed.
We classify the data according to two axes - qualitative versus quantitative and subjective versus objective~\cite{khayat2019validity}.
The second common approach is usage scenarios.
Despite previous calls to distinguish usage scenarios from the more formal case study method~\cite{isenberg2013systematic,schulz2013design},
we find several self-claimed case studies to be actual usage scenarios.
Thirdly,
different from previous surveys,
we find the use of ``interview studies'' where researchers demonstrate the system and cases to the experts and collect feedback through interviews and questionnaires.
In those interview studies,
experts do not work with the system, 
which is different from observational studies.
Fourthly,
there are model experiments to evaluate automated analysis algorithms.
Finally,
we note three controlled experimental studies where the VA system is compared with baselines.
A notable difference is that such studies involve a larger number of participants than observational studies.

\crit{dEva}{lEva}{Evaluations are incomplete and insufficient}
An insight from~\autoref{table:eva} is that evaluating VA systems could require multiple strategies and collect various data for analysis.
For instance,
the surveyed 30 papers, on average, adopt 1.57 strategies and collect 2.47 types of collected data.
Critically, 
the diverse aspects of VA systems often can hardly be tackled with a single evaluation strategy.
For this reason,
evaluations are often criticized for being incomplete and insufficient.
Therefore, it needs to choose appropriate combinations of evaluation methods to validate the argued contributions.


\crit{dEva}{lEva}{Unclear evaluation methods and protocols}
It is important to carry out evaluations using well-established protocols and report the details,
for both without or with human subjects.
For the former,
one should clarify the benchmark datasets and measurements.
For the latter,
the authors should clarify the participants' demographics, user study procedures, data collection, and coding scheme.
Those details enable readers to judge the methodological validity,
promoting reproducibility and comparability.

\crit{dEva}{lEva}{Lacking realism (\eg~real users and datasets)}
Because VA systems are typically driven by real-world problems,
they are expected to be validated under real-world scenarios,
\eg~with real end-users and datasets.
It is a legitimate form of evaluation to present usage scenarios with hypothetical users to describe the workflow and present insights derived from the VA systems.
However, it is considered far more solid and convincing to evaluate with real domain experts~\cite{isenberg2013systematic}.
Some interviewees even argued that ``usage scenarios are not evaluation'' (P2,8).
Besides,
P15 advocated in-the-field evaluations such as longitudinal studies to ``gain more insight into actual use'' and understand ``how the VA systems change the experts' workflow''.

\crit{dEva}{lEva}{Lacking comparison with baseline approaches}
The evaluation sometimes is criticized for not being comparative or controlled.
On the one hand,
interviewees acknowledged the difficulties of carrying out controlled experiments for evaluating VA systems.
Those difficulties have multiple reasons,
such as the difficulties to identify fair baselines in many domains (P1,3,6,12-15),
the lack of benchmark datasets and tasks (P5,6),
``many systems are not open-sourced and thus hard to reproduce'' (P4),
and ``it is unclear how to set users tasks for comparing VA systems'' (P11).
On the other hand,
controlled experiments are a solid and convincing method to convince the benefits of new VA systems.
Thus,
it is worthy to develop, debate, test, and validate comparative evaluation methods,
for example, 
to break complex VA systems into components and conduct ablation studies (P14),
and to compare the final systems with early prototypes (P4).

\crit{dEva}{lEva}{Lacking quantitative evaluation and feedback}
Many evaluation methods such as case studies and expert interviews are qualitative.
Their goal is to maximize the realism of findings,
thereby understanding how the complex systems behave in the field~\cite{carpendale2008evaluating}.
However,
qualitative feedback is subject to experimenter bias, expectancy effects, and other biases in human opinions.
As such,
they are insufficient and thus ``become a weakness in many submissions''.
Quantitative evaluation methods build on measurable variables to interpret the evaluated criteria.
Integrating qualitative and quantitative evaluation could make the evaluation more solid and convincing.

A common method for quantitative feedback is questionnaires.
However,
questionnaires could be less meaningful due to the limited sample size and subjective bias.
P5 and P11 suggested using objective quantitative data such as logs, mouse movement, and eye-tracking data.

\crit{dEva}{lEva}{Analysis insights are suspicious or not new}
It is argued that the purpose of visual analytics is insights~\cite{chen2014visualization}.
Thus, some evaluation methods such as case studies seek to report on analysis workflow and insights,
that is, to demonstrate that users could derive insights from the VA system.
Correspondingly, it degrades the system if the discovered insights are not new or doubtful.
The interviewees suggested evaluating with domain experts to gather feedback about the quality of insights and demonstrate how the insights advance the understandings of the domain problem (P4,7,11,13).
We also note the use of experimental studies to evaluate the quality of insights~\cite{park2021neurocartography},
which might inspire more rigorous approaches for validating the insights.

\crit{dEva}{lEva}{Unclear interpretation of expert feedback}
Feedback from the end-users provides a valuable resource to critically reflect on the VA system.
However, it is not considered a rigorous evaluation by only reporting on the positive feedback that ``might be cherry-picked'' (P14).
Instead, the feedback should be analyzed systematically to provide insightful discussions, especially how the design can be further improved and potentially inform design in related domains.

\subsection{Discussion and Conclusion}

\crit{dDis}{lDis}{Insufficient discussions on limitations and implications}
The discussion should provide critical thoughts on the limitations and implications to inform future work.
However,
it remains challenging to reflect on visualization application research due to the lack of standards~\cite{meyer2018reflection}.
Thus,
many participants found it very challenging to compose reflections (P2,4),
since it requires ``abstracting the experience for the particular VA system to knowledge that generalizes to other domains'' (P12).
Thus,
we encourage thinking broadly about what implications can be beneficial to the visualization community.

\subsection{Others}
We identify several issues that not specific to any of the eight components in VA system manuscript (\autoref{fig:sections}).

\crit{dOthers}{lOthers}{Various presentation issues}
The most frequent one is the presentation issues,
covering varying aspects such as language, grammar, writing organization, and figures (\textbf{I-35}).
In particular,
P6 emphasized that ``figures are particularly important to show professionalism in visualization.''

\crit{dOthers}{lOthers}{Ethics}
Ethics appears as an emerging concern.
It often relates to data privacy, 
especially in sensitive domains such as videos, medicine, and social media (\eg~\cite{zeng2019emoco,zeng2020emotioncues}).
However,
P10 commented that ethics seemed not to gain enough attention in VA system research,
in contrast with empirical research.
She advised researchers to seek approval from the human research ethics committee in prior whenever applicable.


\crit{dOthers}{lOthers}{Open-source}
We observe multiple cases criticizing the research for not making the data and codes public.
During the interviews, we find this issue to be controversial.
Many interviewees agreed that there has been a tendency towards open-source codes in computer science,
and that open-source codes can improve independent validation of the system, promote trust, and accelerate scientific progress.
However,
they pointed out hindrances such as private data and ``many VA systems are prototypes'' (P17).
Those controversies underscore the importance of continuing discussions on the reproducibility of visualization research~\cite{fekete2020exploring}.


\section{Analysis and Implication}
\begin{rev}
In this section, we provide a structured analysis of the interview results.
We first quantitatively analyzed the sample representativeness and the criticisms' specificity to VA systems, followed by a qualitative analysis grouping low-level criticisms to high-level implications and a comparative analysis surfacing pressing issues for the VA community,

\subsection{Sample Statistics}
We analyzed the statistics of our codes in both interview studies.
For the first study, we computed the counts of criticisms ({\fontfamily{lmss}\selectfont{\textbf{C}}}) as a fraction of total coded utterances and per participant.
Each criticism was coded from at least three utterances (Mean: 7.4, SD: 4.6) and two participants (Mean: 6.2, SD: 3.3).
For the second study, we found that every type of criticism was encountered by at least 2 contributors (Mean: 8.2, SD: 3.8) and at least 2 reviewers (Mean: 10.7, SD: 3.9).
No participant reported new types of criticisms not covered in our list.
Therefore, we conclude that our list is reasonably balanced and collectively exhaustive based on our sample.

\begin{figure}[!t]
	\centering
	\includegraphics[width=1\linewidth]{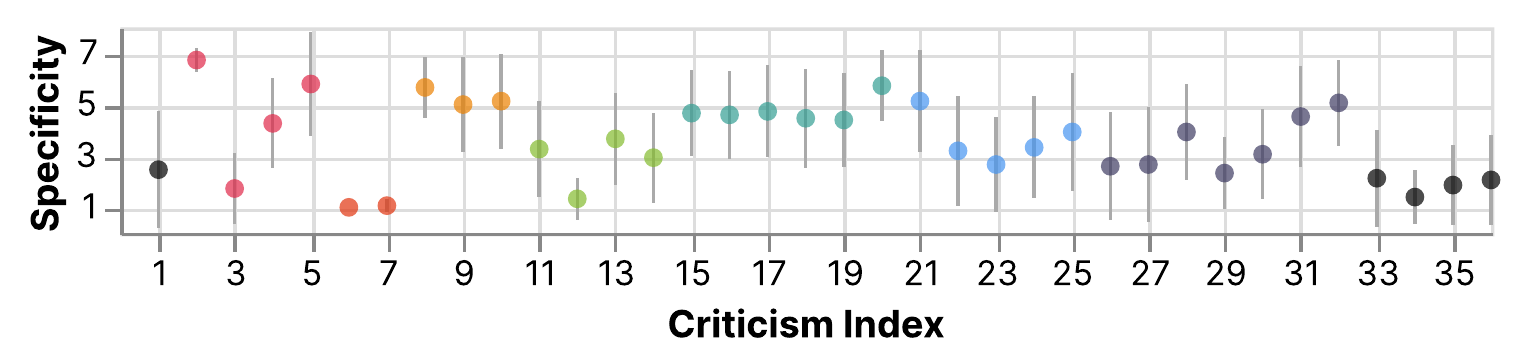}
	\caption{Participants were asked to rate on a 7-Likert scale whether the criticism is not at all (1) or extremely specific (7) to VA systems. This figure shows the average and standard deviation. Color encodes the corresponding components.}
	\label{fig:spec}
	\vspace{-15px}
\end{figure}

\definecolor{mypro}{RGB}{236, 129, 0}
\definecolor{myvisual}{RGB}{52, 159, 148}
\definecolor{mylr}{RGB}{226, 53, 16}
\definecolor{myothers}{RGB}{32, 29, 29}
\definecolor{myEva}{RGB}{63, 60, 99}

\subsection{Specificity to VA Systems}We analyzed interviewees' ratings on the degree that each criticism is specific to VA system research and visualize them in~\autoref{fig:spec}.
On the whole, criticisms on {\color{mypro}Problem Abstraction} and {\color{myvisual}Visual Design} are consistently considered to be more specific,
while criticisms on {\color{mylr}Literature Review} and {\color{myothers}Others} are mostly generic.

On the individual level, relevance to (\ci{2}) and novelty of (\ci{4}) VA systems are rated as the most specific issues, which is not surprising.
More interestingly, over-complicated visual designs (\ci{20}) come third, which suggests complicity has become a vital and shared concern of VA systems.
For example, P1 commented, ``many VA systems do not put into actual application because the systems are too complex to understand and use.''
Other highly-rated issues include unclear target users (\ci{8}), insufficient abstraction from domain to VA problems (\ci{10}),
and lacking workflow overviews (\ci{21}).
Finally, criticisms on analysis insights (\ci{31}) and expert feedback (\ci{32}) stand out from those on {\color{myEva}Evaluation}.
This might imply that researchers tend to consider case studies and expert interviews as distinguishable evaluation methods for VA systems from other research.

\subsection{Implication}
\label{sec:analysis:implication}
Based on the discussion about low-level criticisms in~\autoref{sec:pitfall},
we discuss implications for a high-level question - how to conduct research to defend and improve the research values and rigor of VA systems?

\textbf{Towards developing a criteria checklist.}
The derived low-level criticisms are grouped by corresponding components in the manuscript, which helps structure our discussion.
An alternative is to group them by quality criteria, which could provide a checklist of scientific rigor for researchers to evaluate their own work and conduct more rigorous research.
For example, we could map criticism to the criteria for rigorous visualization research by Lee~\ea~\cite{lee2019beyond},
such as \textit{relevance} (\ci{2,4}), \textit{claims} (\ci{3}),  \textit{originality} (\ci{5,13,16}), \textit{writing} (\ci{6,7,33-34}), \textit{technical soundness} (\ci{22-25}), and \textit{evaluation} (\ci{26-32}).
However, we also find additional ones that can not be readily mapped to above criteria.
For example, \textit{justification} on algorithms (\ci{14}) and visual encodings (\ci{17-19}) helps keep the research informed.
Researchers need to draw on existing research and user feedback to inform the design of VA systems.

\textit{Clarity} appears as another vital concern, requiring researchers to articulate users (\ci{8}), data and algorithms (\ci{11-12}), visual designs (\ci{15}), and systems (\ci{21}).
Its prevailing phenomenon implies that clarity might not be just a writing problem,
but points to structural problems of VA research that there lack documentation standards.
For example, P15 underscored ``inconsistent definitions of goals, requirements, and tasks''.
Thus, continued research is needed to develop standards, guidelines, and common languages to document and communicate VA systems.

\textbf{Extending the scope of VA systems' contributions.}
VA systems are often developed for specific applications on an ad-hoc basis, leading to open questions such as ``what are the values of specific solutions, and do they generalize'' (\eg~\cite{meyer2019criteria,weber2017apply}).
Our report in~\autoref{table:contributions} suggests a wide range of research contributions that can be made by building VA systems,
such as applying them for new domains, characterizing domain problems, novel data analysis and visualization techniques.
In contrast, criticisms can be voiced at the opposites, \eg~failures to characterize domain problems (\ci{9,10}), propose new techniques (\ci{5,13,16}), and demonstrate the ability to address real-world problems (\ci{28}) and derive new insights (\ci{31}). 

Furthermore, generalizability has become a growing concern for VA systems (\ci{24}).
Our analysis of~\ci{4} sheds light on what research contributions of ad-hoc VA systems might generalize to the broader community.
For example, researchers have advocated some actions such as providing transferable reflection (\eg~analyzing visualization design failures and offering suggestions on methodology~\cite{meyer2018reflection}) and contributing open-source toolkits or benchmark datasets~\cite{fekete2020exploring}.
However, \autoref{table:contributions} suggests that such actions remain relatively rare in existing research, therefore requiring more attention from the community to broaden the scope of contributions made by designing, building, deploying, and evaluating ad-hoc VA systems.

\subsection{Comparative Analysis}
Finally, we compare our low-level criticisms with the pitfalls in conducting design studies by Sedlmair~\ea~\cite{sedlmair2012design} to concretize pressing and under-explored issues for the VA community.
Although our criticisms are voiced on written manuscripts and Sedlmair~\ea's pitfalls span the overall study progress,
we find many overlaps.
For example, their ``no need for visualization: problem can be automated'' is related to C2, and ``PF-19: abstraction: too little'' matches with C10.
This suggests strong connections between writing components and design study stages.
However, our criticisms do not cover pitfalls that are usually not externalized in the manuscript, such as ``no real data available'' and ``researcher expertise does not match domain problem''.

More importantly, we find some criticisms that are not completely covered and thus require more attention by the VA community.
First, Sedlmair~\ea discussed two pitfalls in validating and evaluating visualization systems, including ``usage scenario not case study'' and ``liking necessary but not sufficient for validation''.
In contrast, our studies revealed 7 criticisms, which had sparked substantial discussions during our interviews,
\eg~ten participants considered evaluation to be a grand challenge for VA system research.
Second, Sedlmair~\ea listed five pitfalls in the writing stage but did not emphasize clarity.
However, as discussed in~\autoref{sec:analysis:implication}, clarity has emerged as a common problem.
In addition to the lack of documentation standards, explaining complex visualizations has been an open problem~\cite{yang2021explaining,xiong2019curse}. 
This calls for new methods and guidelines to communicate and document VA systems.
\end{rev}

\section{Discussion and Conclusion}
We discuss limitations of this work and future research opportunities to promote the research field forward.

\subsection{Limitation}
We contribute two interview studies to gather common criticisms of VA systems and responses to those criticisms.
We identify two potential threats to the validity of our studies.
First, our interviewees cannot represent the whole community. 
Our recruitment method could introduce bias in our results and harm external validity~\cite{leung2015validity} (\ie~whether the results can generalize to other situations and researcher groups).
To make it clear, our goal is to acquire an initial understanding of common criticisms by interviewing a sufficient number of researchers,
but not to develop exhaustive understandings.
Those studies might require actions from the community, and we hope that our initial results will propel such actions.
Second, we asked interviewees to rephrase peer reviews, which might threaten internal validity~\cite{leung2015validity} (\ie~whether the results are trustworthy).

Our studies are based on qualitative analysis.
There are also potential opportunities to conduct quantitative analysis to gain deeper insights, such as analyzing dependencies among criticisms, and correlating criticisms to other factors like the acceptance result. 
However, such quantitative analysis would require a reasonably large number of original peer-review texts that are beyond our capacity.
Besides, our interviewees are peer researchers and thus not representative of the whole community.
We plan to reach out to a broader range of interviewees, such as researchers out of our field, industrial or governmental practitioners, and the end users of VA systems.


\subsection{Reflection and Future Work}
\begin{rev}
Our studies reveal common criticisms on VA system research. Drawing upon the gathered findings, we reflect on challenging problems for making the research field more rigorous and discuss our perspectives.

\textbf{Constructing knowledge bases to derive general knowledge.}
Research on VA systems has made substantial progress in actual techniques and applications, but considerably less in the theoretical foundation.
Furthermore, the often ad-hoc nature of VA systems has caused concerns about their rigor.
For example, their design and design justifications are often based on feedback from a small group of end-users, leading to questions such as ``are they reliable and representative''.

We argue for the need for constructing knowledge bases of VA systems to theorize about general knowledge.
Our argument is inspired by the recent efforts in building knowledge bases and datasets for visualization research,
such as VisPubData~\cite{isenberg2016vispubdata}, VIS30K~\cite{chen2021vis30k}, and VisImages~\cite{deng2020visimages}.
They offer valuable resources to reflect on visualization research, mine common patterns, and inform future research, \eg~to guide layout designs in multiple-view systems~\cite{chen2020composition} and to summarize frameworks for problem abstraction~\cite{lam2017bridging}.
Similarly, developing a meta-collection of VA systems will enable an inductive approach, that is, to summarize current practices and identify common patterns and anti-patterns to theorize about general knowledge on VA designs.

Addressing those challenges will likely lead to valuable research opportunities.
For instance, there exist different practices and use of terms in problem abstraction, such as design requirements, design rationales, analytical tasks, and visualization tasks.
This issue could prompt researchers to summarize and develop a taxonomy of tasks or requirements in VA systems.
Moreover, the complex visualization designs offer opportunities for us to revisit taxonomies of visualizations,
which could inspire down-streaming applications such as designing declarative languages for VA techniques.

\textbf{Augmenting methodology with a software perspective.}
Research on VA systems has traditionally been rooted in Munzner's nested model~\cite{munzner2009nested} and design study methodology~\cite{sedlmair2012design},
which are greatly informed and influenced by HCI research.
Since they are qualitative and subjective in nature, we often hear researchers asking how to quantify the design and evaluation of VA systems (\eg~\ci{30}).
Due to the interdisciplinary nature of visualization research,
we argue that an inclusive vision to explore alternatives will likely enhance our theoretical and practical underpinnings.
In particular, we argue for a software engineering perspective, as VA systems are software artifacts.

First, software engineering research has proposed many languages such as UML~\cite{booch2005unified} to standardize the disparate systems and represent information such as system structure, behavior, and interaction.
In line with the ever-growing number of VA systems,
we envision that a formal language of VA systems will provide a standard way to document VA systems and conceptual ideas and promote accessibility.

Second, software testing is an objective and quantitative method for validating and verifying software systems.
As evaluating VA systems has been a longstanding challenge,
we ideate that software testing can potentially provide an alternative approach for evaluating VA systems.
For example,
we might run coverage testing~\cite{zhu1997software} to identify components that are rarely used in VA systems,
providing potential opportunities to remove unnecessary components and simplify complex designs.
We hope this perspective could inspire researchers to design and develop other rigorous and feasible approaches to evaluate VA systems.

\textbf{Continuing discussions on the assessment criteria.}
Our study surfaces a common ground of assessment criteria for VA system research.
Those results provide a timely and evidence-based response to the ongoing discussion about standards for rigor.
Not surprisingly, we see that researchers apply different weights to assessment criteria, which are far from reaching a consensus.
For example, there are seemingly tensions between novelty (\ci{5}) and usability (\ci{25}), as many novel VA techniques are ``too complex to use''.
In response, some interviewees argued for changing the current favor of complex visualization designs to embrace actual usability, which needs to be supported and evidenced by longitudinal deployment in the field.

Furthermore, we hear conflicting opinions surrounding some criticisms,
such as whether novel data analysis algorithms are not necessarily well-targeted for the visualization community (\ci{13}),
whether researchers should ask for controlled comparisons between VA systems (\ci{29}), 
whether expert interviews are meaningful evaluation methods (\ci{32}),
and whether open-source should become a norm and even pre-requite (\ci{36}).
Resolving those conflicts will require continued discussions, debates, practices, and research by the community.
We hope our study will inspire and engage researchers to think critically, express opinions, and continue discussions , e.g., at panels, keynotes, and workshops, to move the field forward more rigorously and vibrantly.
\end{rev}

\acknowledgments{
We sincerely thank the participants in our interviews for their kind patience and insightful viewpoints.
This work is partially supported by Hong Kong RGC GRF Grant (No. 16210321).
}

\clearpage

\newpage 
\bibliographystyle{abbrv-doi}

\bibliography{main}




\end{document}